\documentclass[twocolumn,showpacs,showkeys,amsmath,amssymb]{revtex4}

\usepackage{graphicx}
\usepackage{dcolumn}
\usepackage{bm}

\begin{document}

\title{Self-assembled Pt nanowires on Ge(001): Relaxation effects}

\author{U.~Schwingenschl\"ogl and C.~Schuster}
\affiliation{Institut f\"ur Physik, Universit\"at Augsburg, 86135 Augsburg,
Germany}

\date{\today}

\begin{abstract}
Absorption of Pt on the Ge(001) surface results in stable
self-organized Pt nanowires, extending over some hundred nanometers.
Based on band structure calculations within density functional theory
and the generalized gradient approximation, the structural relaxation
of the Ge--Pt surface is investigated. The surface reconstruction
pattern obtained agrees well with findings from scanning tunneling
microscopy. In particular, strong Pt--Pt dimerization is characteristical
for the nanowires. The surface electronic structure is significantly
perturbed due to Ge--Pt interaction, which induces remarkable shifts
of Ge states towards the Fermi energy. As a consequence, the topmost
Ge layers are subject to a metal-insulator transition.
\end{abstract}

\pacs{68.43.Bc, 73.20.-r, 73.20.At}
\keywords{Pt nanowire, self-organization, germanium surface,
density functional theory}

\maketitle

Semiconductor surfaces attract great attention due to a wide field
of possible technological application. In particular, for both the Si and the Ge
surface various effects of self-organization have been reported in the literature.
Self-assembled metallic chains, for example, are formed by Au atoms
on Si(553) \cite{ahn05} and In atoms on Si(111) \cite{hill97}. Electron
motion in Ag films grown on these In chains is restricted to the chain
direction, giving rise to quantized Ag states \cite{nagamura06}. In addition,
ultrathin Ag films on Ge(111) are found to be significantly influenced
by hybridization between the Ag and Ge surface states \cite{tang06}.
For Au growth on Ge(001), a large variety of ordering phenomena can be
realized as a function of the Au coverage and growth temperature \cite{wang0405}.
Adsorption of Pt atoms on the Ge(001) surface induces a self-organization of
well-ordered chain arrays after high-temperature annealing \cite{gurlu03}.
Since these nanowires provide a confining potential, interference between
Ge(001) surface electrons within the nanowire array comes along with
one-dimensional states resembling the energy levels of a quantum particle
in a well \cite{oncel05}. The spontaneously formed Pt chains are
thermodynamically stable, literally defect and kink free, and have lengths
up to some hundred nanometers. Using scanning tunneling microscopy,
Sch\"afer {\it et al.} \cite{schafer06}, however, have shown that their
conduction bands are seriously modified by the interaction
with the Ge substrate. Exact knowledge about the surface electronic states
therefore is mandatory for understanding the self-organization process.

Reconstruction of the clean Ge(001) surface has been subject
to extensive investigations, both from the experimental and theoretical
point of view \cite{zandvliet03}. In particular, neighbouring surface atoms
form asymmetric dimers on top of a slightly relaxed substrate,
saturating one dangling Ge bond per surface atom. Since these dimers
line up along the $\langle110\rangle$ direction, the surface is well
characterized in terms of dimer rows. Further stabilization of the crystal
structure can be reached by means of a distinct buckling of the dimers,
leading to specific reconstruction patterns with respect to the realized
buckling directions. In particular, the room temperature $p(2\times1)$
configuration transforms into a $c(4\times2)$ structure at low temperature,
where the order-disorder phase transition is accompanied by a surface
metal-insulator transition. Scanning tunneling spectroscopy entails
that subtleties of the reconstruction pattern seriously
affect the Ge(001) surface states \cite{gurlu04}. Nevertheless, when
the structural relaxation is carried out carefully, band structure
calculations have shown the capability to perfectly reproduce the
experimental surface density of states (DOS) \cite{prb07}.

\begin{figure}
\includegraphics[width=0.48\textwidth,clip]{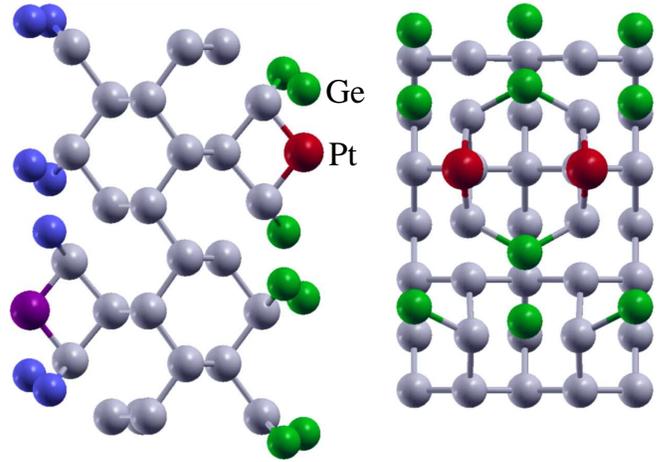}
\caption{(Color online) Crystal structure of the supercell used for
modelling the Ge--Pt surface. Left: Projection along the Pt nanowires
($c$-axis). Right: Projection perpendicular to the surface ($a$-axis),
where the Pt nanowires are oriented from left to right. Pt atoms
are coloured red/violet, surface dimer Ge atoms green/blue, and all
remaining Ge atoms gray.}
\label{fig1}
\end{figure}

A comparative study of the surface reconstructions of diamond, Si, and Ge
has been reported by Kr\"uger and Pollmann \cite{kruger95}, including a
comprehensive review on previous ab initio calculations. While for diamond
a symmetric dimer configuration is established, asymmetric ordering leads
to an energy gain of about 0.1\,eV per dimer in the cases of Si and Ge. The
dimer formation on adsorption of In on Ge(001) has been analyzed by \c{C}akmak
and Srivastava \cite{cakmak04} via first principles total energy calculations.
Their data stress the importance of the very details of the surface electronic
structure for the adsorption mechanism. Expectedly, this likewise applies to the
adsorption of other atoms and molecules, such as alkali metals \cite{xiao06}
or GeH$_4$ \cite{cocoletzi05}, for instance.

Because self-organization of adsorpted atoms relys on strong interaction
with surface states, accounting for the full structural relaxation is a
prerequisite for describing the electronic properties of a covered surface
in an adequate manner. In a first step, the present work therefore aims
at clarifying the reconstruction pattern induced by self-assembled Pt
chains on the Ge(001) surface. Afterwards, we show that the original
Ge(001) surface states are subject to strong modifications due to Ge--Pt
interaction, giving rise to a transition into a metallic regime.

Our first principles calculations are based on a supercell of the cubic
Ge unit cell consisting of one $c(4\times2)$ reconstructed surface array
and extending two unit cells perdendicular to the surface. With
respect to the parent diamond lattice, our tetragonal supercell thus is
given by the lattice vectors $(4,-4,0)$, $(2,2,0)$, and $(0,0,2)$. X-ray
diffraction data of Ferrer {\it et al.} \cite{ferrer95} and first
principles data of Yoshimoto {\it et al.} \cite{yoshimoto00} can be used
for starting the structure optimization, which we first carry out for the
clean Ge(001) surface without Pt chains. Here and in the following, we
apply the generalized gradient approximation within density functional
theory, as implemented in the WIEN2k package \cite{wien2k}. This
full-potential linearized augmented plane wave code has shown great capability
in dealing with structural relaxation at surfaces and interfaces
\cite{uscs1,uscs2}. In particular, the exchange-correlation potential is parametrized
according to the Perdew-Burke-Ernzerhof scheme. We assume convergence of the
relaxation when all the surface forces have decayed, i.e.\ have
reached values below a threshold of 5\,mRyd/$a_B$. To have
reliable results, we represent the charge density by some 460{,}000 plane
waves and use for the Brillouin zone integrations a {\bf k}-mesh comprising
12 points in the irreducible wedge. While Ge $3d$ and Pt $4f$, $5p$ orbitals
are treated as semi-core states, the valence states consist of Ge $4s$,
$4p$, $4d$ and Pt $5d$, $5f$, $6s$, $6p$ orbitals. We have checked convergence
of the band structure results with respect to the thickness of the Ge slab
applied in the calculation.

According to scanning tunneling microscopy data by Gurlu {\it et al.}
\cite{gurlu03}, the formation of Pt chains on Ge(001) is accompanied by the
partial breakup of the Ge surface dimers. These authors have obtained
nanowires only on Pt modified Ge(001) terraces with mixed Ge--Ge and
Ge--Pt dimers. However, as it cannot be excluded that they likewise
can be grown on a pure Ge surface, we have investigated
self-assemblance on the pure and the Pt modified Ge(001) surface. Because
we find qualitatively the same behaviour in both cases, we subsequently
focuss on the pure Ge surface, in order to clearly resolve the effects of
Ge--Pt hybridization under Pt adsorption. To be more specific, the
reconstructed Ge(001) surface consists of Ge troughs, which can accommodate
the nanowires, see the left hand side projection (along the $c$-axis) of
the crystal structure in Fig.\ \ref{fig1}. Due to the occupation of trough
sites by Pt atoms, the upper Ge dimer sites stay unoccupied on one side of
the nanowire, which is illustrated in Fig.\ \ref{fig1} by a projection
perpendicular to the surface (along the $a$-axis). Since only every second Ge trough is occupied, the
distance between neighbouring nanowires along the $b$-axis amounts to
almost exactly 16\,\AA. For obtaining a reasonable starting point for
the structure optimization of the Pt covered Ge(001) surface, we thus
use the unit cell of the clean Ge surface, leave away every
fourth row of upper Ge dimer sites and place Pt sites in the trough next
to this row. As suggested by \cite{gurlu03}, we assume the Pt atoms to be
located in the middle of the troughs (along the $b$-axis) and the Pt--Pt
bond lengths to resemble the distance of adjacant Ge dimers along the row
direction. With respect to the dimer rows, the Pt chain
is shifted by half that distance along the $c$-axis, see Fig.\ \ref{fig1}.

Applying this structure model, surface forces decayed quickly during the
structure optimization. While the Pt sites are subject to strong
relaxation, shifts of Ge sites are small, see Table \ref{tab1}. The lower dimer
sites next to the Pt chain move off the chain axis by 0.05\,\AA, whereas the remaining
upper dimer sites approach it by 0.05\,\AA. Only minor shifts
are found both perpendicular to the surface and in the chain direction.
In contrast, the nanowire develops strong dimerization with alternating Pt--Pt bond
lengths of 3.50\,\AA\ and 4.50\,\AA, confirming the experiments
of Gurlu {\it et al.} \cite{gurlu03}. As a consequence, the
bond length to the nearest upper dimer Ge site decreases to 2.97\,\AA\
and the distance to the nearest lower dimer Ge site grows to 3.83\,\AA. The
shortest Ge--Pt bond lengths of 2.41\,\AA\ and 2.44\,\AA\ appear within
the Ge troughs. Since the Pt chain is not subject to buckling perpendicular
to its axis, the Pt--Pt bond angle maintains a value of almost
180$^\circ$. For convenience, important bond lengths for
the relaxed Ge--Pt surface are summarized in Table \ref{tab2}.
\begin{table}[t]
\begin{tabular}{l|c|c|c}
& $a$-axis & $b$-axis & $c$-axis\\\hline
upper dimer Ge (full row) & $+0.01$\,\AA & $-0.05$\,\AA & $\pm0.00$\,\AA\\
lower dimer Ge (full row) & $-0.01$\,\AA & $+0.05$\,\AA & $\pm0.00$\,\AA\\
lower dimer Ge (partial row) & $-0.01$\,\AA & $-0.05$\,\AA & $\pm0.00$\,\AA
\end{tabular}
\caption{\label{tab1}Atomic shifts of dimer sites next to the Pt chain, with
respect to their positions in the clean Ge surface.}
\end{table}
\begin{table}[t]
\begin{tabular}{l|c}
& bond length\\\hline
upper dimer Ge -- Pt & 2.97\,\AA\\
lower dimer Ge -- Pt & 3.83\,\AA\\
trough Ge -- Pt & 2.41\,\AA, 2.44\,\AA\\
Pt -- Pt (along $b$-axis) & 16.0\,\AA \\
Pt -- Pt (along $c$-axis) & 3.50\,\AA, 4.50\,\AA
\end{tabular}
\caption{\label{tab2}Selected bond lengths in the relaxed Ge--Pt surface,
as obtained from the structure optimization.}
\end{table}

\begin{figure}
\includegraphics[width=0.42\textwidth,clip]{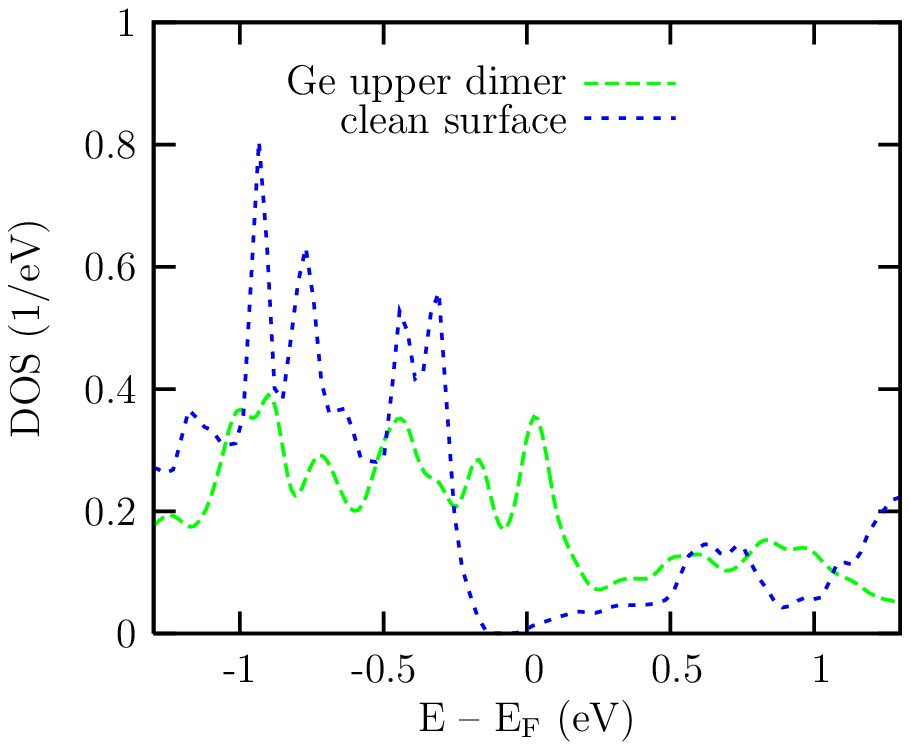}\\
\includegraphics[width=0.42\textwidth,clip]{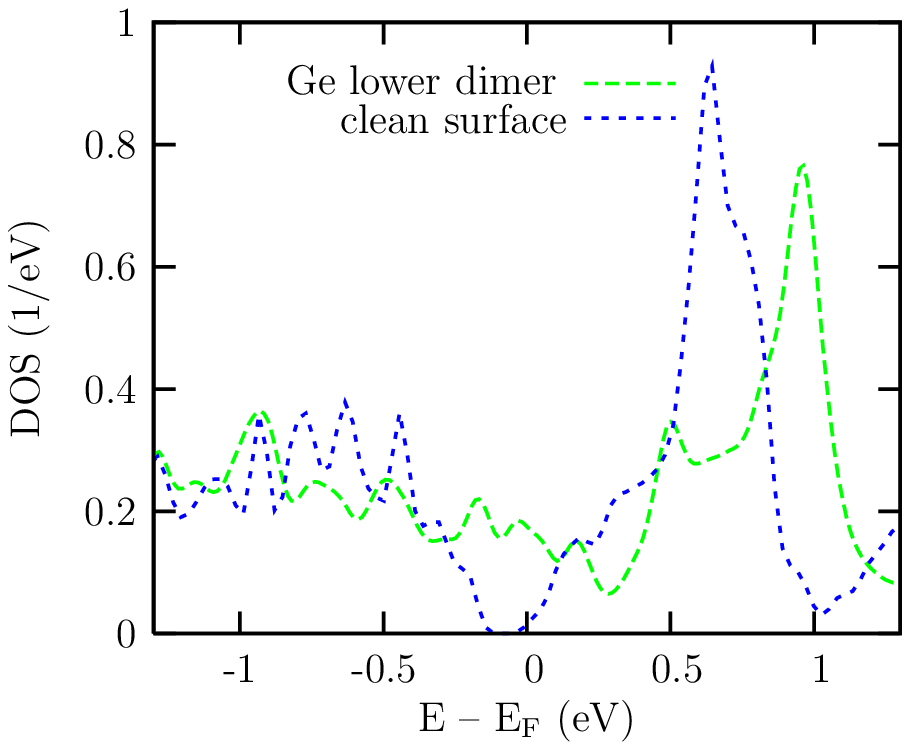}\\
\includegraphics[width=0.42\textwidth,clip]{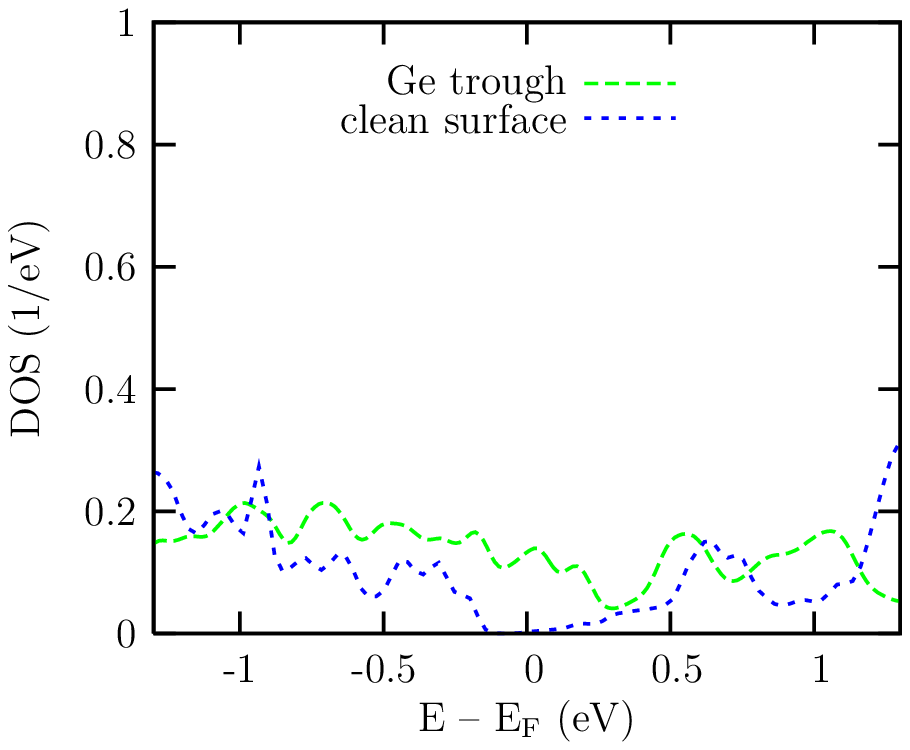}
\caption{(Color online) Partial Ge densities of states for an upper
dimer site (top), lower dimer site (middle), and trough site (bottom).
The electronic structure of the Ge-Pt surface is compared to a clean
Ge surface. Additional surface states are formed close to the Fermi energy.}
\label{fig2}
\end{figure}

Having clarified the structural details of the Pt covered Ge(001) surface,
we now turn to the relaxation effects on the Ge surface states. Fig.\ \ref{fig2}
compares partial Ge densities of states (DOS) as resulting from our band
structure calculation for the relaxed Ge--Pt surface to data of the clean reconstructed
surface \cite{prb07}. The three panels shown refer to three characteristic atomic sites:
the upper and lower dimer site as well as a typical trough site. For the clean
surface, electronic states tracing back to the upper Ge dimer atom are found
mainly below the Fermi level. In contrast, the DOS of the lower Ge dimer atom
has its maximum above the Fermi level, which reflects a remarkable amount of
charge transfer from the lower to the upper dimer site. Importantly, for all
Ge sites spectral weight is missing at the Fermi energy, which
agrees perfectly with the prediction of an insulating surface state, see \cite{gurlu04}
and the references in this paper.

Accounting for the Pt chains, the shapes of the DOS curves in Fig.\
\ref{fig2} change dramatically. For the upper Ge dimer site,
weight is lost at energies below $-0.2$\,eV and reappears near the Fermi
energy, hence closing the insulating energy gap. Alterations of the
unoccupied states are much less pronounced. For the lower Ge dimer site,
the situation is virtually inverted. While strong changes affect the
unoccupied states, again shifting weight to the Fermi energy, the occupied
states mainly maintain their original shape. In a first step, we
interpret these facts in terms of perturbed Ge--Ge dimer bonds.
Charge transfer from the lower to the upper dimer atom, in principle,
results in a filled D$\rm_{up}$ and an empty D$\rm_{down}$ surface band,
which opens the energy gap. Due to surface strain induced by the
Pt chains, the Ge dimers are forced to give up their ideal geometry and
charge transfer is reduced as a consequence. The D$\rm_{up}$
band hence is partially depleted while the D$\rm_{down}$ band is filled
accordingly. In addition to the variation of the dimer bonds, the surface
states likewise are influenced by strong hybridization with the Pt
conduction bands. This becomes obvious on inspection of the trough Ge
atoms, which are not involved in dimer bonding, see the bottom panel of
Fig.\ \ref{fig2}. In the energy range from $-0.8$\,eV to 0.3\,eV
the DOS grows seriously for the Pt covered surface, particularly
inducing a metallic state. Since each of the surface Ge atoms is observed to
develop conduction states, the coaction of structural relaxation and
Ge--Pt hybridization effectively causes an metal-insulator transition
of the entire surface.

In conclusion, we have studied the relaxation and electronic structure of
the $c(4\times2)$ reconstructed Ge(001) surface covered with self-organized
Pt nanowires. The surface geometry and electronic states have been discussed
using results from band structure calculations based on density functional
theory. We have carefully analyzed the surface reconstruction pattern induced
by adsorption of Pt atoms, confirming a strong tendency towards Pt--Pt
dimerization with intrachain bond lengths of 3.50\,\AA\ and 4.50\,\AA. In
addition, we have succeeded in establishing a comprehensive structure model
for a Pt covered Ge(001) surface. By a fully relaxed crystal structure,
alterations of the Ge surface states due to interaction with the Pt chains
could be investigated. Our results are evident as concerns the existence of
a surface phase transition into a metallic state. We attribute this transition
to two cooperative effects: (1) A shift of surface Ge states towards the Fermi
energy due to a modified bonding within the Ge dimers and (2) considerable
hybridization of the Ge states and the metallic Pt states. The metallicity
of the topmost Ge layers is expected to significantly affect the transport
properties of the Pt nanowires. Moreover, the observed interplay between a strong
Pt--Pt dimerization and Ge--Pt hybridization agrees well with the data of recent
experiments by scanning tunneling microscopy \cite{schafer06}.

\subsection*{Acknowledgement}
We thank U.\ Eckern and V.\ Eyert for helpfull discussions, and
the Deutsche Forschungsgemeinschaft for financial support (SFB 484).

\end{document}